\documentclass{article}

    \PassOptionsToPackage{numbers, compress}{natbib}

\usepackage[preprint]{neurips_2022}

\usepackage{wrapfig}
\usepackage[utf8]{inputenc} 
\usepackage[T1]{fontenc}    
\usepackage{hyperref}       
\usepackage{url}            
\usepackage{booktabs}       
\usepackage{amsfonts}       
\usepackage{nicefrac}       
\usepackage{microtype}      
\usepackage{xcolor}         
\usepackage{caption}
\usepackage{subcaption}
\usepackage{bbm}
\usepackage{times}
\usepackage{soul}
\usepackage{graphicx}
\usepackage{amsmath}
\usepackage{amsthm}
\usepackage{algorithm}
\usepackage{amsmath}
\usepackage{amssymb}
\usepackage{amsthm}
\usepackage{bm}
\usepackage{algpseudocode}

\newtheorem{defi}{Definition}

\newenvironment{proof-sketch}{\noindent{\textit{Sketch of Proof.}}\hspace*{1em}}{\qed\bigskip}

\title{Two-Stage Neural Contextual Bandits for \\Personalised News Recommendation}
\author{%
  Mengyan Zhang \thanks{Email: \textit{mengyan.zhang@anu.edu.au}}\\
  Australian National University\\
  Data61, CSIRO\\
  \And
  Thanh Nguyen-Tang\\
  Deakin University\\
  \And
  Fangzhao Wu\\
  Microsoft Research Asia\\
  \And
  Zhenyu He\\
  University of Electronic Science\\
   and Technology of China
  \And
  Xing Xie\\
  Microsoft Research Asia\\
  \And
  Cheng Soon Ong\\
  Data61, CSIRO\\
  Australian National University\\
}

\begin{document}

\maketitle

\begin{abstract}
We consider the problem of personalised news recommendation where each user consumes
news in a sequential fashion.
Existing personalised news recommendation methods focus on exploiting user interests and ignores exploration in recommendation, which leads to biased feedback loops and hurt recommendation quality in the long term.
We build on contextual bandits recommendation strategies which naturally address the exploitation-exploration trade-off.
The main challenges are the computational efficiency for exploring the large-scale item space and utilising the deep representations with uncertainty. 
We propose a two-stage hierarchical topic-news deep contextual bandits framework to efficiently learn user preferences when there are many news items.
We use deep learning representations for users and news, and generalise the neural upper confidence bound (UCB) policies to generalised additive UCB and bilinear UCB.
Empirical results on a large-scale news recommendation dataset show that our proposed policies are efficient and outperform the baseline bandit policies.
\end{abstract}

\section{Introduction}

Online platforms for news rely on effective and efficient personalised news recommendation \citep{wu2021personalized_survey}.
The recommender system faces the \textit{exploitation-exploration dilemma}, where one can exploit by recommending items that the users like the most so far, or one can also explore by recommending items that users have not browsed before but may potentially like \citep{li_contextual-bandit_2010}. 
Focusing on exploitation tends to create a pernicious feedback loop, which amplifies biases and raises the so-called \textit{filter bubbles} or \textit{echo chamber} \cite{jiang_filter_bubble_2019},
where the exposure of items is narrowed by such a self-reinforcing pattern.

Contextual bandits are designed to address the exploitation-exploration dilemma and have been proposed used to mitigate the feedback loop effect \cite{chen_biasrec_survey_2020, li_contextual-bandit_2010} by user interest and item popularity exploration. 
One can formalise the online recommendation problem as a sequential decision-making under uncertainty, where given some contextual information, an agent (the recommender system) selects one or more arms (the news items) from all possible choices according to a policy (recommendation strategy), with the goal of designing a policy which maximises the cumulative rewards (user clicks). 

There are two main challenges on applying contextual bandits algorithms in news recommendations. 
First, the recommendations need to be scalable for the large news spaces with millions of news items, which requires the bandit algorithms to learn efficiently when there are many arms (news items). 
Second, contextual bandits algorithms need to utilise good representations of both news and users. 
The state-of-the-art news recommender systems utilise deep neural networks (DNN) with two-tower structures (user and news encoders) \cite{wu_nrms_2019}.
How to combine contextual bandits models with such DNN models with valid uncertainty estimations remains an open problem.
We review related work which addresses each challenge respectively in Section \ref{sec: related work}. 

\begin{figure}
    \centering
    \includegraphics[scale=0.8]{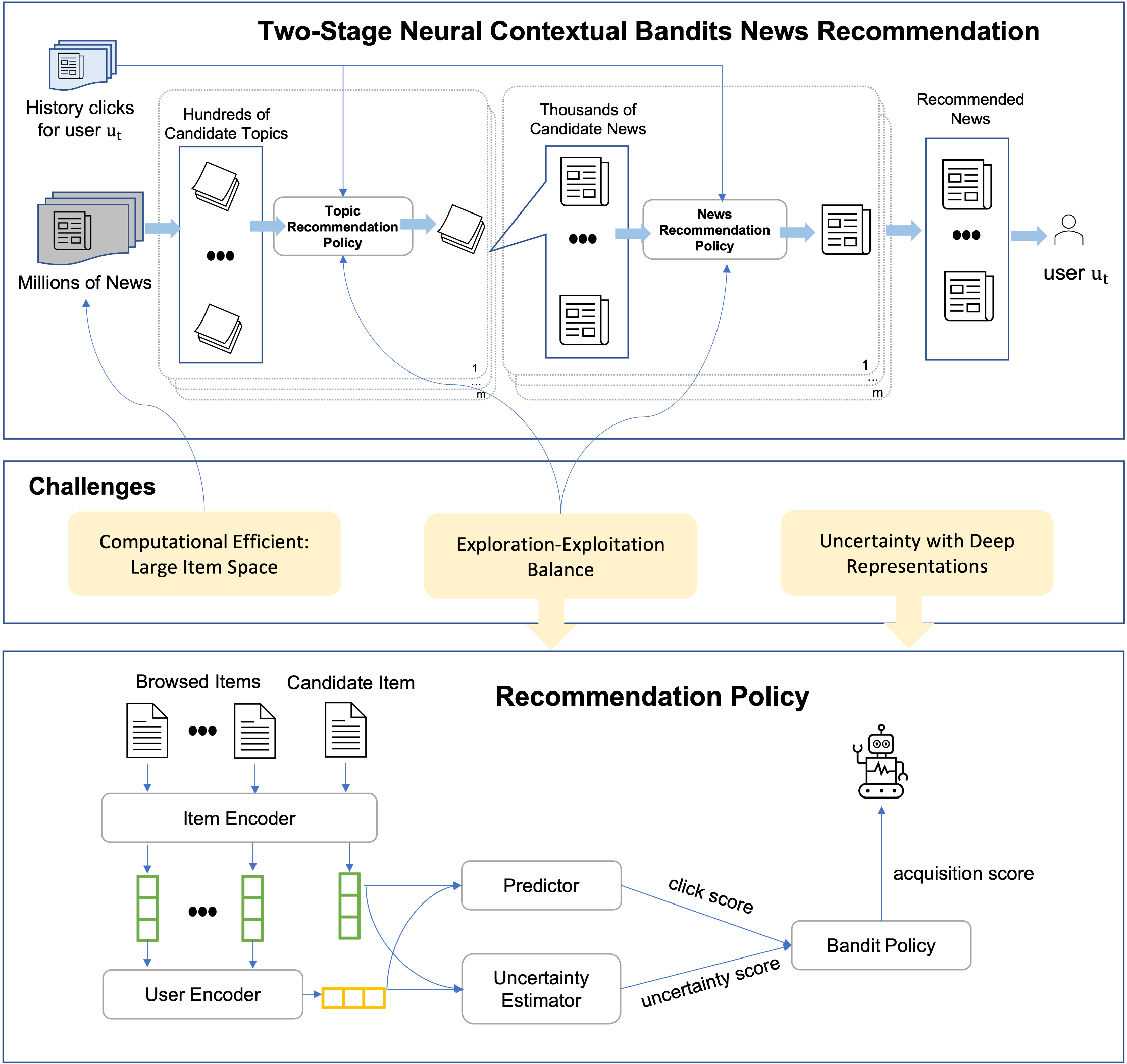}
    \caption{Two-stage neural contextual bandits framework for news recommendation. We address the large item space using a two-stage hierarchy of topic and news items. 
    }
    \label{fig:figure 1}
\end{figure}

We propose a two-stage neural contextual bandits framework to address the above challenges, and illustrate the mapping in Figure \ref{fig:figure 1}.
We consider a hierarchical topic-news model, 
where for each of the recommendations for one user in one iteration, we recommend topics first and then select an item from the recommended topics. 
For each stage, 
we utilise the state-of-the-art two-tower deep model NRMS \cite{wu_nrms_2019} to generate topic, news and user representations.
We propose shared neural generalised additive and bilinear upper confidence bound (UCB) policies, and extend existing neural contextual bandits approaches like Monte-Carlo dropout \cite{gal_dropout_2016} UCB, neural-linucb \cite{deepshowdown_2018,xu_deep_shallow_2020} to our framework as baselines.
We evaluate our proposed framework empirically on a large-scale news recommendation dataset MIND \cite{wu_mind_2020} and compare our proposed policies with baseline approaches.
To our knowledge, we are the first \st{work} to apply two-stage neural contextual bandits framework to address above challenges. 

Our \textbf{contributions} are 
1) We propose a hierarchical two-stage neural contextual bandits framework for  user interest exploration in news recommendation,
where in the first stage we dynamically construct topics.
2) We propose  shared neural generalised additive and bilinear upper confidence bound policies to make use of the deep representation of contextual information.
3) We conduct experiments to simulate the user interest exploration and compare with baseline policies on a large-scale real-world news recommendation dataset.


\section{Problem Setting and Challenges}
\label{sec:problem-setting} 

\paragraph{Personalised News Recommendation}
We consider a news recommender system that sequentially recommends personalised news to users, with the goal of maximising cumulative clicks for all users.
The recommender system learns from the interaction history with the users, and for any given coming user, the system displays several news selected from the candidate news set.
Then the user will react as either click or non-click and the system uses this as the feedback to learn the preference of users. 
This task is challenging since the candidate news set is in the millions and dynamically changes over time.
In addition, there are a large number of cold users (i.e. users that do not have any history) and the user interest can shift over time \citep{wu2021personalized_survey}.
How to design such a recommendation strategy can be formulated as a sequential decision-making problem, studied in the field of contextual bandits \citep{li_contextual-bandit_2010, song_hucb_2021}. 

\paragraph{Cumulative Reward}
We first introduce the general bandit problem formulation.
A news recommender system is regarded as an \textit{agent}, news items are \textit{arms} (choices), and the user and/or item embedding form the \textit{context}. 
At each iteration $t = 1,\dots, N$,
given user $u_t$ and candidate arm set $\mathcal{A}_t$, one can generate the item embedding $\bm{x}_i \in \mathbb{R}^{d_1}$ for all $i \in \mathcal{A}_t$, and the user embedding $\bm{z}_{u_t} \in \mathbb{R}^{d_2}$ as context. 
In the following, we will drop subscript $t$ for $u_t$ when there is no ambiguity. 
The agent recommends $m \geq 1$ news items $\mathcal{S}_{rec}$ according to a \textit{policy} $\pi$ given the context.
Then the agent receives the feedback $\{y_{t,1}, \dots, y_{t,m}\}$, where $y_{t,i} \in \{0,1\}$ indicating whether the user clicks the item $i$ or not at iteration $t$. 
The \textit{reward} is defined as $y_t = \sum_{i = 1}^{m} \mathbb{I}\{y_{t,i} = 1\}$.
The goal is to design a policy to minimise the expected cumulative regret (Definition \ref{defi: expected cumulative regret}), which is equivalent to maximise the expected cumulative rewards \citep{li_contextual-bandit_2010, song_hucb_2021}.  
Since in recommender systems the optimal rewards are usually unknown, we focus on maximising the cumulative rewards in this work.
\begin{defi}
\label{defi: expected cumulative regret} 
For a total iteration $N$, the expected cumulative rewards are $\mathbb{E} \left[\sum_{t=1}^N y_t \right]$.
Let the optimal reward for user $u_t$ as $y_t^\ast$, the expected cumulative regret is defined as $\mathbb{E} \left[\sum_{t=1}^N (y_t^\ast - y_t) \right].$
\end{defi}





\paragraph{Contextual Bandits Policies}
Upper Confidence Bound (UCB) are one type of classical bandits policies proposed to address the exploration-exploitation dilemma and proven to have sublinear regret bound \cite{auer_finite-time}. 
The idea is the picking the arm with the highest UCB \textit{acquisition score}, which capture the upper confidence bound for predictions in high probability.
At iteration $t$, the UCB \textit{acquisition function} for a pair user-item $(u,i)$ follows
\begin{align}
\label{equ: ucb policy}
    \alpha^{UCB} (u, i) := \hat{y}_{u, i} + \beta \sigma_{u, i},
\end{align}
where $\hat{y}_{u, i}$ is the click prediction, $\sigma_{u, i}$ is the uncertainty of predictions, and $\beta$ is a hyperparameter balancing the exploitation and exploration. 
\citet{li_contextual-bandit_2010} popularised the LinUCB contextual bandits approach on news recommendation tasks, where the expected reward of item $i$ and user $u$ is assumed to be linear in terms of the contextual feature $\bm{c}_{u,i} \in \mathbb{R}^d$. 
\citet{xu_deep_shallow_2020, deepshowdown_2018} studied neural linear models, where the representation of contextual information is learnt by neural networks, which further improves the performance.
\citet{filippi_GLMUCB_2010} extended the LinUCB policy to the Generalised Linear Model s.t.
$\mathbb{E}[y_{u,i}|\bm{c}_{u,i}] = \rho(\bm{c}_{u,i}^T \bm{\theta}^\ast_{u})$,
where $\rho: \mathbb{R} \rightarrow \mathbb{R}$ is the inverse link function, $\bm{\theta}^\ast_{u} \in \mathbb{R}^d$ is the unknown coefficient.
When $\rho(\bm{x}) = \bm{x}$, the problem is reduced to linear bandits. 
Define the \textit{design matrix} $D_{u} \in \mathbb{R}^{n_{u} \times d}$ at iteration $t$, where each row contains sample interacted with user $u$.
With $M_{u} = D_{u}^T D_{u} + \bm{I}_d$ and estimated coefficient $\hat{\bm{\theta}}_{u}$, the GLM-UCB acquisition function follows
\begin{align}
\label{equ: glmucb}
    \alpha^{GLM-UCB}(u,i) := \rho(\bm{c}_{u,i}^T \hat{\bm{\theta}}_{u}) + \beta 
    {\|\bm{c}_{u,i}\|}_{M_{u}^{-1}}.
\end{align}
In this paper, we consider GLM-UCB policy as a base policy. Since user feedback is binary, 
we use the sigmoid function, i.e. we set $\rho(x) = \exp(x)/(1+\exp(x))$, which is the inverse link function of a Bernoulli distribution \citep*{nelder1972generalized}.

\paragraph{Two-tower User and Item Representation Learning}
We consider the state-of-the-art news recommendation model NRMS \cite{wu_nrms_2019} as our base model for predictor, which is a two-tower neural network model with multi-head self-attention. 
At each stage and time step $t$, we maintain two modules: 1) The item encoder $f^n_t$, which takes item information in (e.g. for topic recommendation: topic id, topic name; for news recommendation: news id, title, abstract, etc;) and outputs a news embedding $\bm{x} \in \mathbb{R}^{d_1}$, and 
2) the user encoder $f^u_t$, which takes the browsed news embedding of user $i$ in and outputs a user embedding $\bm{z} \in \mathbb{R}^{d_2}$.
The user and news embedding are treated as context, and the arms (choices) are the candidate news available to be recommended to the coming user.

\subsection{Challenges}
\label{sec:challenges}

\textbf{Computational Efficiency: Large Item Space: }
Large scale commercial recommender system has millions of dynamically generated items.
Calculating the acquisition scores for all candidate news can be computationally expensive.
We address this by proposing a two-stage framework by selecting topics first in Section \ref{sec: Two-Stage Deep Recommendation Framework}.
In terms of uncertainty inference, while Bayesian models provide distribution predictions and have shown good performance in bandits tasks, it is computationally expensive to maintain Bayesian neural models and updates for large scale systems.
Two-tower recommendation model is popular in practical usage due to its efficient inference. 
We consider the upper confidence bound (UCB) based policies and combine it with the two-tower deep learning framework with the additional generalised linear model to inference uncertainties.

\textbf{Exploration-Exploitation: Uncertainty with Deep Representation: }
Greedily recommending news to users according to predictors learnt by user clicks may lead to feedback loop bias and suboptimal recommendations. 
Thus, an appropriate level of online explorations can guide the system to dynamically track user interests and contributes to optimal recommendations. State-of-the-art news recommender systems make use of deep neural networks to learn news and user representations. 
How to make use of the power of deep representation and calculating uncertainties (i.e. confidence interval for predictions) is the key point of efficient exploration. 
We propose two exploration policies to address this in \ref{sec: proposed policies}.
We further propose to dynamically form the topic set according to the bandits acquisition score, which avoids biased exploration due to unbalanced topics. 


\section{Two-Stage Deep Recommendation Framework}
\label{sec: Two-Stage Deep Recommendation Framework}




\begin{algorithm}[t!]
  \caption{Two-stage Exploration Framework}
  \label{alg: Two-stage Exploration Framework}
  \textbf{Input:} number of items to be recommended $m$, number of simulation $N$, number of topics $q$, 
  $N$ users $\{u_1, \dots, u_N\}$,
  set of topics $\mathcal{V} = \{v_1, \dots, v_j, \dots, v_{q}\}$, each topic is associated with a set of news $\mathcal{S}_{v_j}$,
  topic acquisition function $\alpha_1$, news acquisition function $\alpha_2$.  
  \label{alg: Two stage exploration}
  \begin{algorithmic}[1]
    \For{$t = 1$ to $N$}
        \State{\underline{\textbf{Stage One: Topic recommendation}}}
        \State{Score each topic $\alpha_1(u_t, v_j)$ for all $v_j \in \mathcal{V}$}
        \label{algo-line: begin stage one}
        \State{Sort topic $v_{(1)}, \dots, v_{(q)}$ according to topic acquisition scores in non-increasing order s.t. $\alpha_1(u_t, v_{(1)}) \geq \dots \geq \alpha_1(u_t, v_{(q)})$}
        \State{Pick sets of topics as $\mathcal{S}^1 = \{v_{(1)}\}, \dots, \mathcal{S}^m = \{v_{(m)}\}$}
        \State Dynamic topic set reconstruction: expand $\mathcal{S}^1, \dots, \mathcal{S}^m$ with high-score topics.
        \label{algo-line: end stage one}
        \State{\underline{\textbf{Stage Two: News recommendation}}}
        \For{j = 1 to m}
        \label{algo-line: begin stage two}
            \State{Scoring each item $\alpha_2(u_t, i)$ for all $i \in \mathcal{S}_{v}$, for all $v \in \mathcal{S}^j$.}
            \State{Add the item $i^\ast$ with the highest score $\alpha_2(u_t, i^\ast)$ to recommendation set $\mathcal{S}_{rec}$.
            }
        \EndFor
        \label{algo-line: end stage two}
        
        \State{Show $\mathcal{S}_{rec}$ to user $u$ and get feedback $[y_{u,a}]_{a \in \mathcal{S}_{rec}} = $ oracle(u, $\mathcal{S}_{rec}$)}
        \State{Update topic and item models with rewards $[y_{u,a}]_{a \in \mathcal{S}_{rec}}$.}
    \EndFor
  \end{algorithmic}
\end{algorithm}
 \vspace{-2mm}

Recall our goal is to sequentially recommend $m \geq 1$ news items to users in a large scale recommender system.
To reduce the computational complexity of whole-space news exploration, we consider a two-stage exploration framework in Algorithm \ref{alg: Two-stage Exploration Framework}.
We call each of the $m$ recommendation as \textit{recommendation slot}.
In stage one (line \ref{algo-line: begin stage one}-\ref{algo-line: end stage one}), we recommend a set of topics for each recommendation slot.
Each topic is treated as an arm, and we decide which topics can be recommended by the topic acquisition function $\alpha_1$. For example, one can use the UCB acquisition function defined in E.q. (\ref{equ: ucb policy}) or (\ref{equ: glmucb}) as $\alpha_1$.
For each recommendation slot, we initialise the set of recommended topics with the top $m$ acquisition scores respectively.
Then in line \ref{algo-line: end stage one}, we dynamically expand each of the topic set with the remaining high-score topics.
In stage two (line \ref{algo-line: begin stage two}-\ref{algo-line: end stage two}), we select the most promising news item (according to the bandit acquisition function $\alpha_2$) for each of the expanded set of topics chosen in stage one.
The acquisition functions in Algorithm \ref{alg: Two-stage Exploration Framework} used to recommend topics and news can follow any contextual bandits policies.
We introduce the baselines and proposed policies we used in this work in Section \ref{sec: baseline policies} and \ref{sec: proposed policies}, which are summarised in Table \ref{table: Summary of UCB policies.}.

\paragraph{Dynamic Topic Set Reconstruction} 
It is common that the sizes of the first stage arms are imbalanced. For example, if one clusters items based on similarity, it is highly likely the clusters will end up to be imbalanced. 
In our application, news topics are highly imbalanced, ranging from size of 1 up to 15,000 (number of news per topics). 
We propose to address the imbalanced topics issue by dynamically reconstructing the set of topics corresponding to each arm according to topic acquisition scores in each iteration. 
After the dynamic topic set reconstruction, each topic set has at least $p$ candidate news items.
The main idea of forming topic sets is to include the topics with high bandits acquisition scores, which means these topics are either potential good exploitation or exploration for user interest. 
Furthermore, we also want to allocate topics with high acquisition scores into different topic sets, so that topics with high scores will have more chance to be selected.
We initialise each topic set with the top $m$ scoring topics  $\{ \mathcal{S}^1, \dots, \mathcal{S}^m \}$.
Then until all topic sets have at least $p$ news items, we add the topic with the highest topic acquisition score in the remaining topics to each of the $m$ topic set  in sequential order. 
We illustrate the detailed description in Algorithm \ref{alg: Dynamic Topic Reconstruction} in the Appendix.


Once the agent collects $m$ recommended items (one news item per each of $m$ recommended topics), those $m$ items will be shown to the user and the agent will get user feedback, which is $m$ binary scores indicating click or non-click for each recommended item. 
The topic and news neural models are updated according to the feedback every $l_t$ and $l_n$ (pre-defined hyperparameters) iterations respectively.
The coefficients of generalised linear models are updated every iteration if applicable.

\subsection{Baseline Neural Contextual Bandits Policies: Exploration}
\label{sec: baseline policies}

\begin{table}[t!]
    \centering
    \small
     \caption{Summary of UCB policies. Recall $\bm{x}_{i}, \bm{z}_{u}$ are the item and user representation, $\hat{\bm{\theta}}_{u}, \hat{\bm{\theta}}_{x}, \hat{\bm{\theta}}_{z}, \hat{\bm{\theta}}$ are coefficients in generalised linear models, with respect to each user, all items, all users, and all user-item pairs respectively. 
    $f^u_{t-1}, f^n_{t-1}$ are user, item encoders up to iteration $t-1$. $\rho$ is the sigmoid function. $Y^n_{u,i}$ is a list of predictions via Monte-Carlo dropout, 
    $\tilde{\gamma} = 1 - \gamma \in (0,1).$ 
    Note we allow biases inside the parameters, i.e. $\bm{x}_i$ is argumented as $(x_{i,1}, \dots, x_{i,d}, 1)$.}
    \label{table: Summary of UCB policies.}
    \vspace{2mm}
    \begin{tabular}{ccccc}
    \hline
    \begin{tabular}[c]{@{}c@{}}Policy\\ Name\end{tabular} & \begin{tabular}[c]{@{}c@{}}Context \end{tabular} & \begin{tabular}[c]{@{}c@{}}Coefficients\\ Parameters\end{tabular} & \begin{tabular}[c]{@{}c@{}}Predicted \\ Rewards\end{tabular} & \begin{tabular}[c]{@{}c@{}}Predicted \\ Uncertainty\end{tabular} \\ \hline \hline
    GLM &  $\bm{x}_{i}$   &  $\hat{\bm{\theta}}_{u}$  & $\rho(\bm{x}_{i}^T \hat{\bm{\theta}}_{u})$  &  ${\|\bm{x}_{i}\|}_{M_{u}^{-1}}$  \\ 
    N-GLM &  $\bm{x}_{i}$   &  $\hat{\bm{\theta}}_{u}, f^n_{t-1}$  & $\rho(\bm{x}_{i}^T \hat{\bm{\theta}}_{u})$  &  ${\|\bm{x}_{i}\|}_{M_{u}^{-1}}$  \\ 
    N-Dropout   &  $\bm{x}_{i}, \bm{z}_{u}$  &   $f^u_{t-1}, f^n_{t-1}$    &     mean($Y^n_{u,i}$)      &    std($Y^n_{u,i}$)       \\ 
    S-N-GALM   &  $\bm{x}_{i}, \bm{z}_{u}$  &   $f^u_{t-1}, f^n_{t-1}, \hat{\bm{\theta}}_x, \hat{\bm{\theta}}_z$  &    $\rho(\gamma \bm{x}_{i}^T \hat{\bm{\theta}}_x +  \tilde{\gamma}{\hat{\bm{\theta}}_z}^T \bm{z}_{u})$  &   $\gamma {\|\bm{x}_i\|}_{A_i^{-1}} + \tilde{\gamma} {\|\bm{z}_u\|}_{A_u^{-1}}$   \\ 
    S-N-GBLM   &   $\bm{x}_{i}, \bm{z}_{u}$    &    $f^u_{t-1}, f^n_{t-1}, \bm{\hat{\theta}}$  & $\rho(\bm{x}_{i}^T \hat{\bm{\theta}} \bm{z}_{u})$    &  ${\|vec(\bm{x}_{i}  \bm{z}_{u}^T)\|}_{W_t^{-1}}$  \\ \hline
    \end{tabular}
    \vspace{-5mm}
\end{table}

\looseness=-1
Recent work have studied neural contextual bandits algorithms theoretically \cite{zhou_neural_2020, xu_deep_shallow_2020} and empirically \cite{deepshowdown_2018}, according to those we adapt two most popular algorithms into our framework \cite{collier_deepCB_2018, xu_deep_shallow_2020} (see below).
\paragraph{Neural Dropout UCB (N-Dropout-UCB)}
As studied by \citet{gal_dropout_2016}, the uncertainty of predictions can be approximated by dropout applied to a neural network with arbitrary depth and non-linearity. Dropout can be viewed as performing approximate variational inference, with a variational family that is a discrete distribution over the value of the parameters and zero.
Dropout UCB policies follow this principle, where for user-item pair $(u, i)$, one can predict the click scores with Monte-Carlo dropout enabled,   $Y^n_{u,i} = [\hat{y}^1_{u,i}, \dots, \hat{y}^n_{u,i}]$, where $\hat{y}_{u,i} = f^u_{t-1}(u) ^T f^n_{t-1}(i)$.
Then using the mean of the predictions $\bar{y}_{u, i}$ as central tendency and the standard deviation $\sigma_{u, i}$ as the uncertainty, one can follow UCB policy defined in E.q. (\ref{equ: ucb policy}).

\paragraph{Neural Generalised Linear UCB (N-GLM-UCB)} To utilise the representation power of DNNs and the exploration ability from linear bandits,  
Neural-Linear \citep{deepshowdown_2018, xu_deep_shallow_2020} learns contextual embedding from DNNs and use it as input of a linear model.
Since our reward is binary, we extend neural LinUCB \cite{xu_deep_shallow_2020} to
neural generalised linear UCB, where we first get the deep contextual embedding learnt from NRMS model, and then follow the same acquisition function as in E.q. (\ref{equ: glmucb}). 
Applying existing neural contextual bandits algorithms directly on recommender systems may be computationally expensive or lead to suboptimal performance. 
For example, uncertainties inferred from Monte-Carlo can have high variance \cite{deepshowdown_2018}.
Also, learning coefficients for each arm in neural-linear models is unrealistic, since one needs enough samples for each of the millions of news items.  
In our simulation, the number of users is much smaller than the news items, hence we learn coefficients per user. 
From our experiment in Table \ref{table: one stage ctr}, we observe that performance still drop when the number of users increases. 


\subsection{Proposed Policies: Additive and Bilinear UCB}
\label{sec: proposed policies}


We consider \textit{shared} bandits models where the parameters are shared by all pairs of users and (or) news items. 
Coefficient sharing across entities can make the model learned more efficient and more generalisable.
One also needs to design how to capture both the item and user embedding in the contextual information. We propose the \textit{generalised additive linear} or \textit{generalised bilinear} models to handle this.
Recall $\bm{x}_i \in \mathbb{R}^{d_1}$ as item $i$ representation and $\bm{z}_u \in \mathbb{R}^{d_2}$ as user $u$ representation. 

\paragraph*{Shared Neural Generalised Additive Linear UCB (S-N-GALM-UCB)} 
We consider an additive linear model, where the item-related coefficient $\bm{ \theta}^\ast_x$ and user-related coefficient ${\bm{\theta}^\ast_z}$ are modelled separately, i.e. 
$
    \mathbb{E}[y_{u,i}|\bm{x}_{i},\bm{z}_{u}] =  \rho(\gamma \bm{x}_{i}^T \bm{ \theta}^\ast_x + \tilde{\gamma} {\bm{\theta}^\ast_z}^T \bm{z}_{u}),
$
where $\gamma$ is a hyperparameter, $\tilde{\gamma} = 1 - \gamma$. 
\begin{align}
\label{equ: Hybrid Neural DualUCB}
    \alpha^{S-N-GALM-UCB}(u,i) := \rho(\gamma \bm{x}_{i}^T \hat{\bm{\theta}}_x +  \tilde{\gamma} {\hat{\bm{\theta}}_z}^T \bm{z}_{u})+ \beta 
   (\gamma {\|\bm{x}_i\|}_{A_i^{-1}} + \tilde{\gamma} {\|\bm{z}_u\|}_{A_u^{-1}}),
\end{align}
where $A_i = D_i^T D_i + \bm{I}_{d_1}$, 
with $D_i \in \mathbb{R}^{n_i \times d_1}$ be a design matrix at iteration $t$, where each row contains item representations that user $u$ that has been observed up to iteration $t$;
$A_u = D_u^T D_u + \bm{I}_{d_2}$, 
with $D_u \in \mathbb{R}^{n_u \times d_2}$ be a design matrix at iteration $t$, 
where each row contains user representations that item $i$ has been recommended to up to iteration $t$.
In this way, the additive model handles the user and item uncertainties separately.


\paragraph*{Shared Neural Generalised Bilinear UCB (S-N-GBLM-UCB)} Inspired by the Bilinear UCB algorithm (rank $r$ Oracle UCB) proposed by \citet{jang_bilinearBandits_2021}, we consider a Generalised bilinear model, where we assume
$
\mathbb{E}[y_{u,i}|\bm{x}_{i},\bm{z}_{u}] = \rho(\bm{x}_{i}^T \bm{\theta}^\ast \bm{z}_{u}),
$ 
with the coefficient $\bm{\theta}^\ast$ shared by all user-item pairs.
\begin{align}
    \alpha^{S-N-GBLM-UCB}(u,i) := \rho(\bm{x}_{i}^T \hat{\bm{\theta}} \bm{z}_{u}) + \beta 
    {\|vec(\bm{x}_{i}  \bm{z}_{u}^T)\|}_{W_t^{-1}},
\end{align}
where $W_t = W_0 + \sum_{s=1}^{t-1} vec(\bm{x}_{i_s}  \bm{z}_{u_s}^T) vec(\bm{x}_{i_s}  \bm{z}_{u_s}^T)^T \in \mathbb{R}^{d_1d_2 \times d_1d_2}$, and $W_0 = \mathrm{I}_{d_1d_2}$.
Computing the confidence interval might be computationally costly due to the inverse of a potentially large design matrix. Different from \citet{jang_bilinearBandits_2021}, instead recommending a pair of arms $(u,i)$, 
we consider the item $i$ as arm to be recommended, and user $u$ as side information instead of an arm. 
The two-tower model in recommender system is naturally expressed in terms of bilinear structure. 

    A bilinear bandit can be reinterpreted in the form of linear bandits \cite{jang_bilinearBandits_2021}, 
    $\bm{x}_{i}^{T} \bm{\theta}^{*} \bm{z}_{u} =\left\langle\operatorname{vec}\left(\bm{x}_{i} \bm{z}_{u}^{\top}\right), \operatorname{vec}\left(\bm{\theta}^{*}\right)\right\rangle$.
    So linear bandits policies can be applied on bilinear bandits problem with regret upper bound $\tilde{\mathcal{O}}(\sqrt{d_1^2 d_2^2 T})$, 
    where $\tilde{\mathcal{O}}$ ignores polylogarithmic factors in $T$.
    However naive linear bandit approaches cannot fully utilise the characteristics of the parameters space.
    The bilinear policy \citep{jang_bilinearBandits_2021} shows the regret upper bound $\tilde{\mathcal{O}}(\sqrt{d_1 d_2 d r T})$, with $d = \max \left(d_{1}, d_{2}\right) \text { and } r=\operatorname{rank}\left(\bm{\theta}^{*}\right)$.
\subsection{Related Work}
\label{sec: related work}


\paragraph{Hierarchical Exploration}
To address large item spaces, hierarchical search is employed. For two-stage bandits work, \citet{hron_exploration_2020,hron_component_2021} studied the effect of exploration in both two stages with linear bandits algorithms and Mixture-of-Experts nominators. 
\citet{ma_off-policy_2020} proposed off-policy policy-gradient two stage approaches, however, without explicit two-stage exploration. 
There is also a branch of related work considering hierarchical exploration. 
\citet{wang2018HMAB,song_hucb_2021} explored on a pre-constructed tree of items in MAB or linear bandits setting. 
\citet{zhang_ConUCB_2020} utilises key-terms to organise items into subsets and relies on occasional conversational feedback from users. 
As far as we know, no existing work studies two-stage exploration with deep contextual bandits. 

\paragraph{Neural Contextual Bandits} 
Contextual bandits with deep models have been used as a popular approach since it utilise good representations.
\citet{deepshowdown_2018} conducted a comprehensive experiment on deep contextual bandits algorithms based on Thompson sampling, including dropout, neural-linear and bootstrapped methods. 
Recently, there are work applying deep contextual bandits to recommender system. 
\citet{collier_deepCB_2018} proposed a Thompson sampling algorithm based on inference time Concrete Dropout \citep{gal_concretedropout_2017} with learnable dropout rate,
and applied this approach on marketing optimisation problems at HubSpot. 
\citet{guo_dbb_2020} studied deep Bayesian bandits with a bootstrapped model with multiple heads and dropout units, which was evaluated offline and online in Twitter's ad recommendation. 
\citet{wang22_context_uncertainty} added representation uncertainty for embedding to further encourage explore items whose embedding have not been sufficiently learned based on recurrent neural network models. 

Theoretically, \citet{zhou_neural_2020} proposed NeuralUCB and proved a sublinear regret bound, 
followed which \citet{gu_batched_2021} studied the case where the parameters of DNN only update at the end of batches. 
\citet{xu_deep_shallow_2020} proposed Neural-LinUCB to make the use of deep representation from deep neural networks and shallow exploration with a linear UCB model, and provided a sublinear regret bound. 
\citet{zhu_sau_2021} proposed sample average uncertainty frequentist exploration, which only depends on value predictions on each action and is computationally efficient. 

To the best of our knowledge, among those utilised the power of deep representation from existing network structures in online recommender system with bandits feedback, no existing work addressed the generalised bilinear model for exploration, which suits the two-tower recommender system naturally; and no work has addressed the hierarchical exploration, which can increase the computational efficiency and is important to the practical use in a large-scale recommender system.

\section{Experiments}
\label{sec: Experiments}
We conduct experiments on a large-scale news recommendation dataset, i.e. MIND \cite{wu_mind_2020}, 
which was collected from the user behaviour logs of Microsoft News.~\footnote{\url{https://microsoftnews.msn.com}} The MIND dataset contains 1,000,000 users, 161,013 news, 285 topics and 24,155,470 samples, which is split to train, validation and test data for machine learning algorithm usage. 

We simulate the sequential recommendation based on MIND dataset. The experiments run in $T$ independent trials. For each trial $\tau \in [1,T]$, we randomly select a set of users $\mathcal{U}_{\tau}$ from the whole user set as the candidate user dataset from trial $\tau$.
We randomly select $\epsilon \%$ of samples $\mathcal{S}_{known}$ from the MIND-train dataset as known data to the bandit models and can be used to pre-train the parameters of bandits neural model NRMS.
Note we have removed the samples of the users in $\mathcal{U}_{\tau}$ from $\mathcal{S}_{known}$ for each trial $\tau$ to avoid leak information.
This simulates the case where in recommender system we have collected some history clicks for other users and we would like to recommend news to new users sequentially and learn their interests. 
In each iteration $t$ of  the total $N$ simulation iterations within 
each trial $\tau$, we randomly sample a user $u_t \in \mathcal{U}_{\tau}$ to simulate the way user $u_t$ randomly shows up to the recommender system. 

To illustrate how the computational complexity of algorithms influence the performance, we follow \citet{song_hucb_2021} and introduce the \textit{computational budget} $b = 5000$, which constraints the maximum number of acquisition score over arms one can compute before conducting the recommendation. The computational budget is set to evaluate the computational efficiency of algorithms and is meaningful for practical applications like large-scale recommender system. 
For one-stage algorithms, we randomly sample $b$ news from the whole news set for the candidate news set of iteration $t$; for two-stage algorithms, we first query all topics then use the left budget to explore the items.


We evaluate the performance by the cumulative rewards as defined in Definition \ref{defi: expected cumulative regret}. 
To make the score more comparable between different number of recommendations, we further define the click-through-rate (CTR) inside a batch of $m$ recommendations at iteration $t$ fo each trial $\tau$ as $\textrm{CTR}^{\tau}_{t} = \frac{1}{m} \sum_{i = 1}^m \mathbb{I} \{y^{\tau}_{t} = 1\}$.
Then we evaluate the performance of bandits policies by the cumulative CTR over $N$ iterations  $\sum_{t=1}^N \textrm{CTR}^{\tau}_{t}$.
We report the mean and standard deviation of the cumulative reward or CTR over $T$ trials.

Evaluation of contextual bandits algorithms on recommendation system is challenging.
On the one hand, deploying algorithms in live recommender systems can be logistically and economically expensive. 
On the other hand, directly evaluating on logged the sparse recommendation data would constrain the exploration effects. 
We consider the \textit{off-policy evaluation} approach, and build a \textit{user-choice simulator} to simulate user feedback for any given news.
We train the simulator on the logged data (MIND-train) and evaluate different methods with the same simulator. 

\subsection{Main Results}
\label{sec: main results}

\begin{table}[t] 
    \centering
    \caption{Cumulative CTR for one stage policies with different number of users. We recommend 5 news for each user and simulate the experiment with 2,000 iterations, 5 trials. In policy names, ``S'' means shared parameters, and ``N'' means using neural contextual information from the NRMS.}
    \label{table: one stage ctr}
    \vspace{0.1cm}
    \begin{tabular}{cccc} 
        \hline
        \textbf{Policy \textbackslash{} \# User} & \textbf{10}          & \textbf{100}         & \textbf{1,000}         \\ 
        \hline
        \hline
        Random                          & 320 $\pm$ 4            & 320 $\pm$ 2            & 320 $\pm$ 2             \\ 
        GLM                          & 442 $\pm$ 4            & 340 $\pm$ 4            & 320 $\pm$ 6             \\ 
        
        N-GLM                        & 1,140 $\pm$ 39         & 522 $\pm$ 58           & 341 $\pm$ 11            \\ 
        N-Greedy                       & 1,188 $\pm$72           & 1,244 $\pm$ 38          & 1,282 $\pm$ 46           \\ 
        N-Dropout                    & 1,198 $\pm$ 41          & 1,256 $\pm$ 34          & 1,286 $\pm$ 44           \\ 
        \hline
        S-N-GALM                       & \textbf{1,538 $\pm$ 20} & \textbf{1,522 $\pm$ 20} & \textbf{1,540 $\pm$ 19}  \\ 
        S-N-GBLM                      & 1,402 $\pm$ 42          & 1,366 $\pm$ 21          & 1,362 $\pm$ 38           \\
        \hline
        \end{tabular}  
    \end{table}
    
Directly learning from the logged data suffers selection bias and affects the simulator learned from it. 
Follow \cite{huang_keeping_2020,song_hucb_2021}, we adopted a standard method used for off-policy evaluation of bandit algorithms, the Itermediate Bias Mitigation Step via the \textit{Inverse Propensity Score} (IPS) simulator \cite{imbens2015causal}, which re-weigh the training samples by the inverse propensity score. 
In particular, we learn the IPS from logged data via logistic regression \cite{schnabel_IPS_2016}.
We then convert the predicted scores $\hat{y} \in [0,1]$ from simulator to binary rewards $\{0,1\}$ by picking a threshold in order to serve the bandits simulation.
We pick the threshold with the largest f-score on validation dataset, where f-score is defined as $2  (\text{precision} \cdot \text{recall})/(\text{precision} + \text{recall})$.
To simulate the stochastic rewards, we flip the reward with probability $\varepsilon = 0.1$.

We first evaluate the bandits policies for one-stage exploration to illustrate the improvement with utilising deep representations and the effectiveness of our proposed policies. 
We evaluated all policies with 2,000 iterations and 5 trials and show the cumulative CTR with one standard deviation in Table \ref{table: one stage ctr}. 
For \textit{Random} policy, we recommend news uniformly at random from the sampled news set $\mathcal{N}_t$.
For \textit{GLM-UCB} policy shown in E.q. (\ref{equ: glmucb}), the news item representation uses GloVe \citep{pennington2014glove} vectors of the news titles, while in \textit{N-GLM-UCB}, we use the NRMS model. In both policies, we learn $\hat{\bm{\theta}}_u$ with collected data for each user $u$.
\textit{N-Greedy} refers to the policy recommending arms greedily with the NRMS model predictions and is the baseline for neural network based policies. 
For \textit{N-Dropout-UCB}, we infer 5 times with dropout enabled. For all UCB based algorithm, we set the exploitation-exploration balance parameter $\beta = 0.1$.
Results are shown in Table~\ref{table: one stage ctr}.

    \begin{table}[t]
        \caption{Cumulative CTR for policies with different number of recommendation each iteration. We select 100 users and simulate the experiment with 2,000 iterations, 5 trials. The prefix ``2-'' indicates two-stage policies. }
\label{table: two stage ctr}
\vspace{0.1cm}
        \centering
        \begin{tabular}{cccc} 
\hline
\textbf{Policy \textbackslash{} \# Recs} & \textbf{1}          & \textbf{5}           & \textbf{10}           \\ 
\hline
\hline
Random                                  & 298 $\pm$ 1           & 320 $\pm$ 2            & 300 $\pm$ 3             \\ 
N-Greedy                                & 418 $\pm$ 224         & 1,244 $\pm$ 38          & 1,364 $\pm$ 32           \\ 
N-Dropout                            & 428 $\pm$ 222         & 1,256 $\pm$ 34          & 1,368 $\pm$30            \\ 
S-N-GALM                                & 424 $\pm$ 6           & 1,522 $\pm$ 20          & 1,506 $\pm$ 42           \\ 
S-N-GBLM                                & 422 $\pm$ 6           & 1,366 $\pm$ 21          & 1,443 $\pm$ 6            \\ 
\hline
2-Random                               & 228 $\pm$ 2           & 252 $\pm$ 4            & 250 $\pm$ 2             \\ 
2-N-Greedy                             & 426 $\pm$
  435       & 1,450 $\pm$ 28          & 1,315 $\pm$ 28           \\ 
2-N-Dropout                         & 438 $\pm$ 452         & 1,470 $\pm$ 38          & 1,326 $\pm$ 28           \\ 
\hline
2-S-N-GALM                             & \textbf{444 $\pm$ 10} & \textbf{1,674 $\pm$ 45} & \textbf{1,578 $\pm$ 25}  \\
2-S-N-GBLM                             & 428 $\pm$ 8           & 1,655 $\pm$ 23          & 1,556 $\pm$ 22           \\ 
\hline
\end{tabular}
\end{table}

For the two-stage experiments, we used 100 users, up to 2,000 iterations and 5 trials. 
We tested recommendation size $\{1,5,10\}$ in each iteration for each user and show results in Table~\ref{table: two stage ctr}.
We select the one-stage policies in Table \ref{table: one stage ctr} that perform well (beyond $1,000$ cumulative CTR) under $100$ users, namely N-Greedy, N-DropoutUCB, S-N-GALM-UCB and S-N-GBLM-UCB, and test their performance with additional topic-stage exploration. 
For two-stage policies, topic and item parts follow the same policy.
The last four rows follow the Algorithm \ref{alg: Two-stage Exploration Framework}.
For two-stage Random policy, we first select topics uniformly random from all topics and then randomly select news from the selected topics.

\vspace{-4mm}
\subsection{Observations and Interpretations}
\begin{figure}[t!]
    \centering
    \begin{subfigure}[b]{0.48\textwidth}
        \centering
        \caption{}
        \label{subfig: Improvement over random policy}
        \includegraphics[scale=0.42]{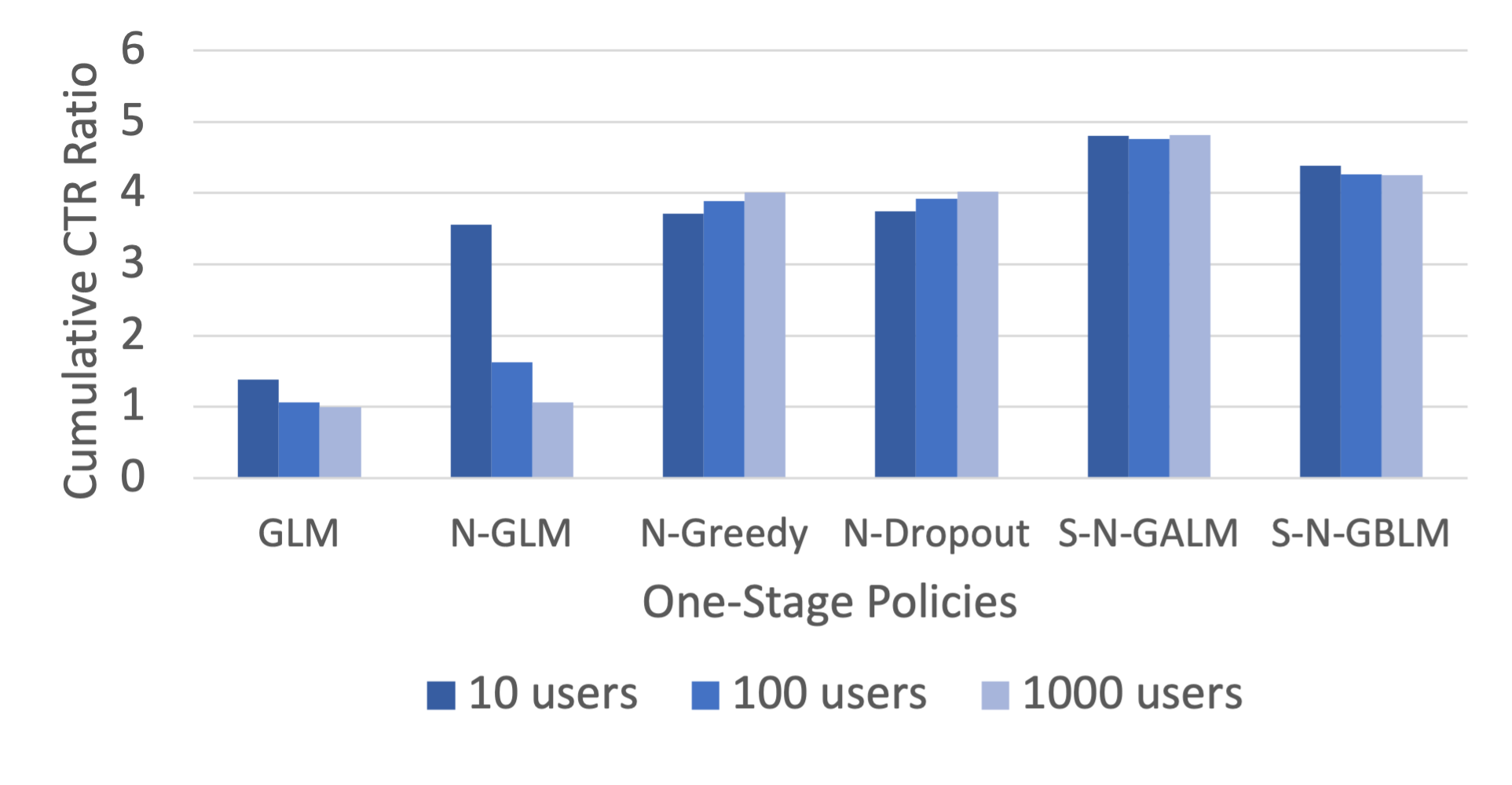}
    \end{subfigure}
    \begin{subfigure}[b]{0.48\textwidth}
        \centering
        \caption{}
        \label{subfig: Influence of number of users}
        \includegraphics[scale=0.42]{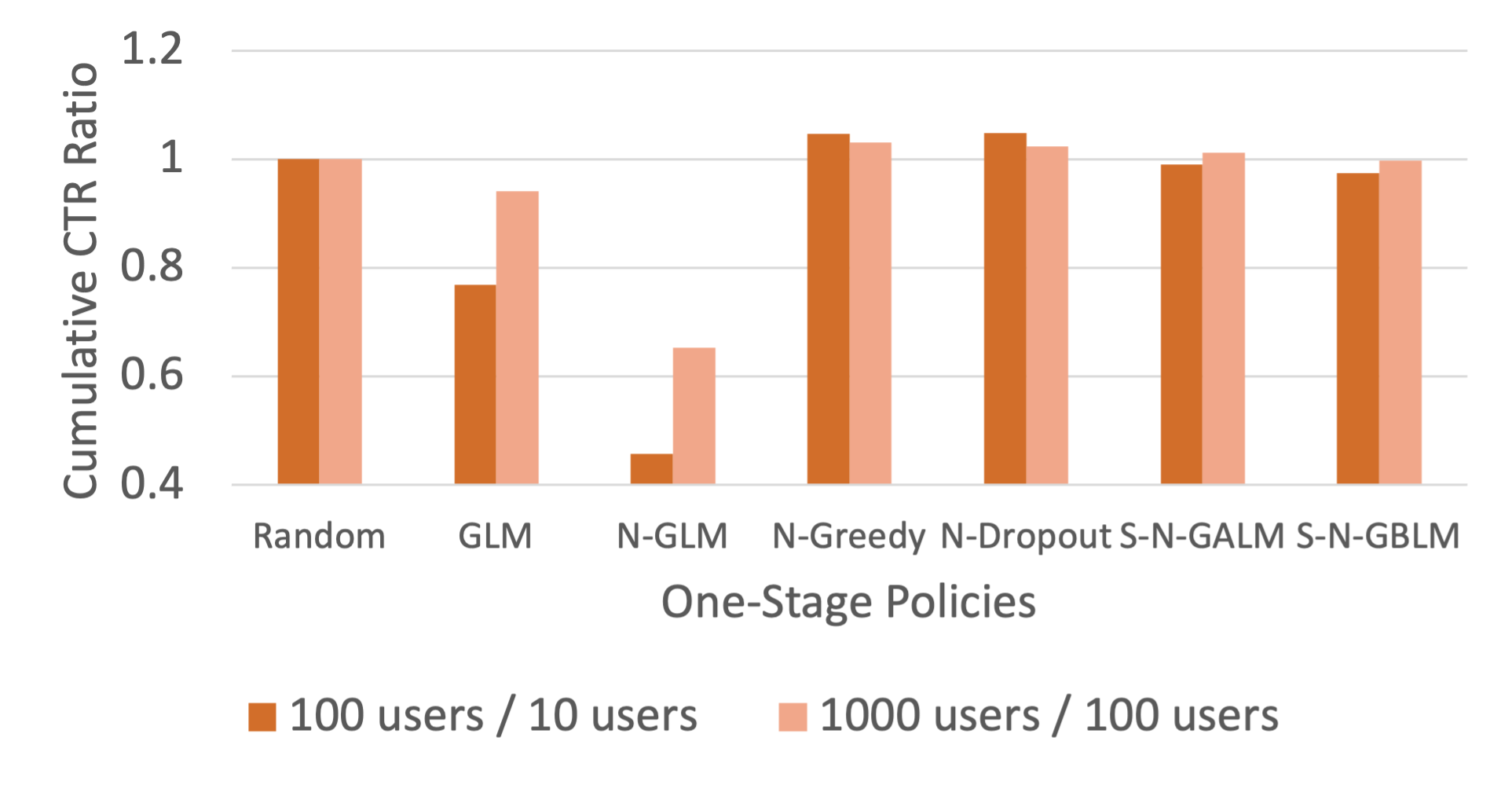}
    \end{subfigure}
    \begin{subfigure}[b]{0.48\textwidth}
        \centering
        \caption{}
        \label{subfig: Improvement of two-stage design}
        \includegraphics[scale=0.42]{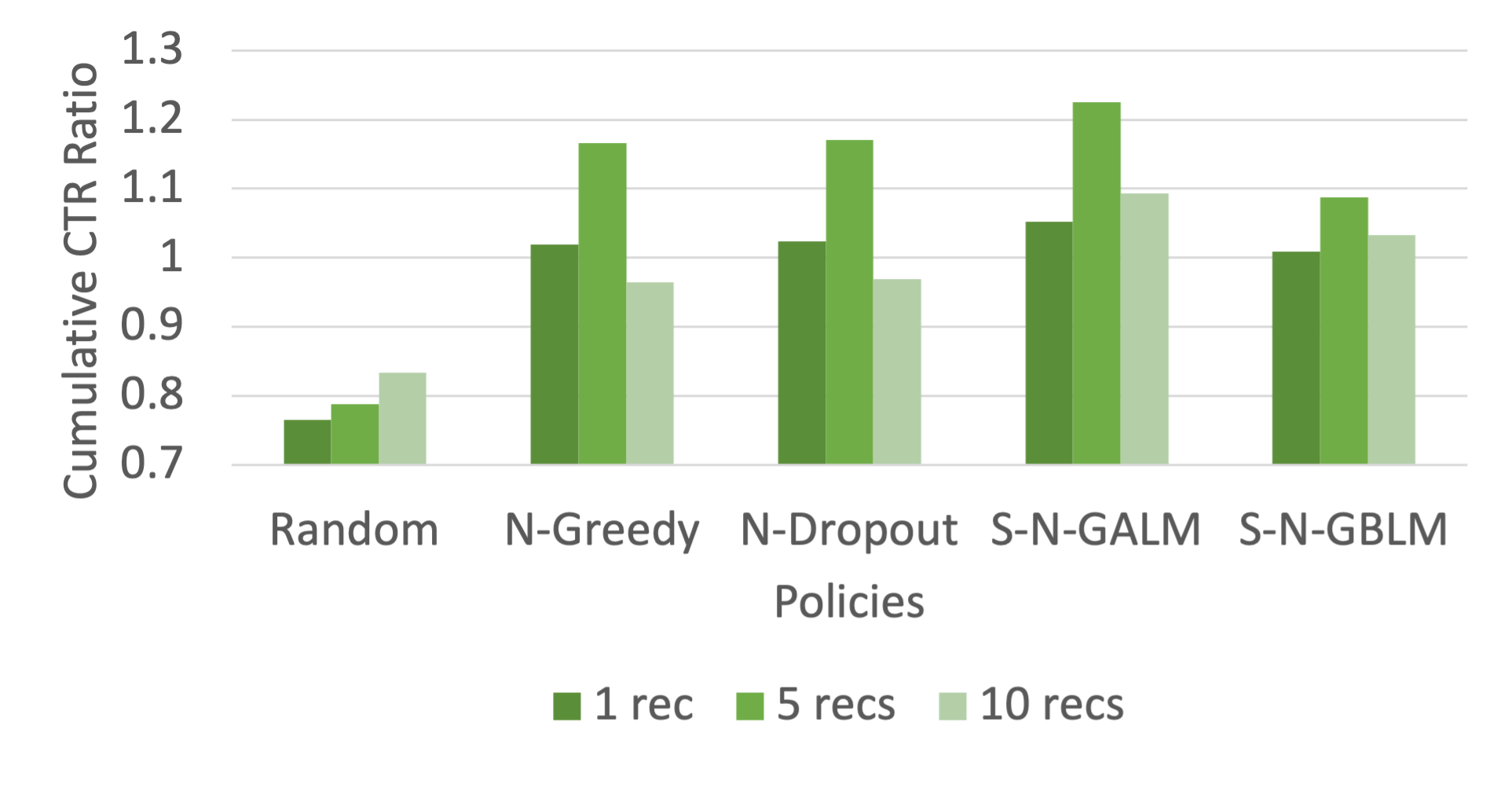}
    \end{subfigure}
    \begin{subfigure}[b]{0.48\textwidth}
        \centering
        \caption{}
        \label{subfig: Ablation study for dynamic topic comparison}
        \includegraphics[scale=0.42]{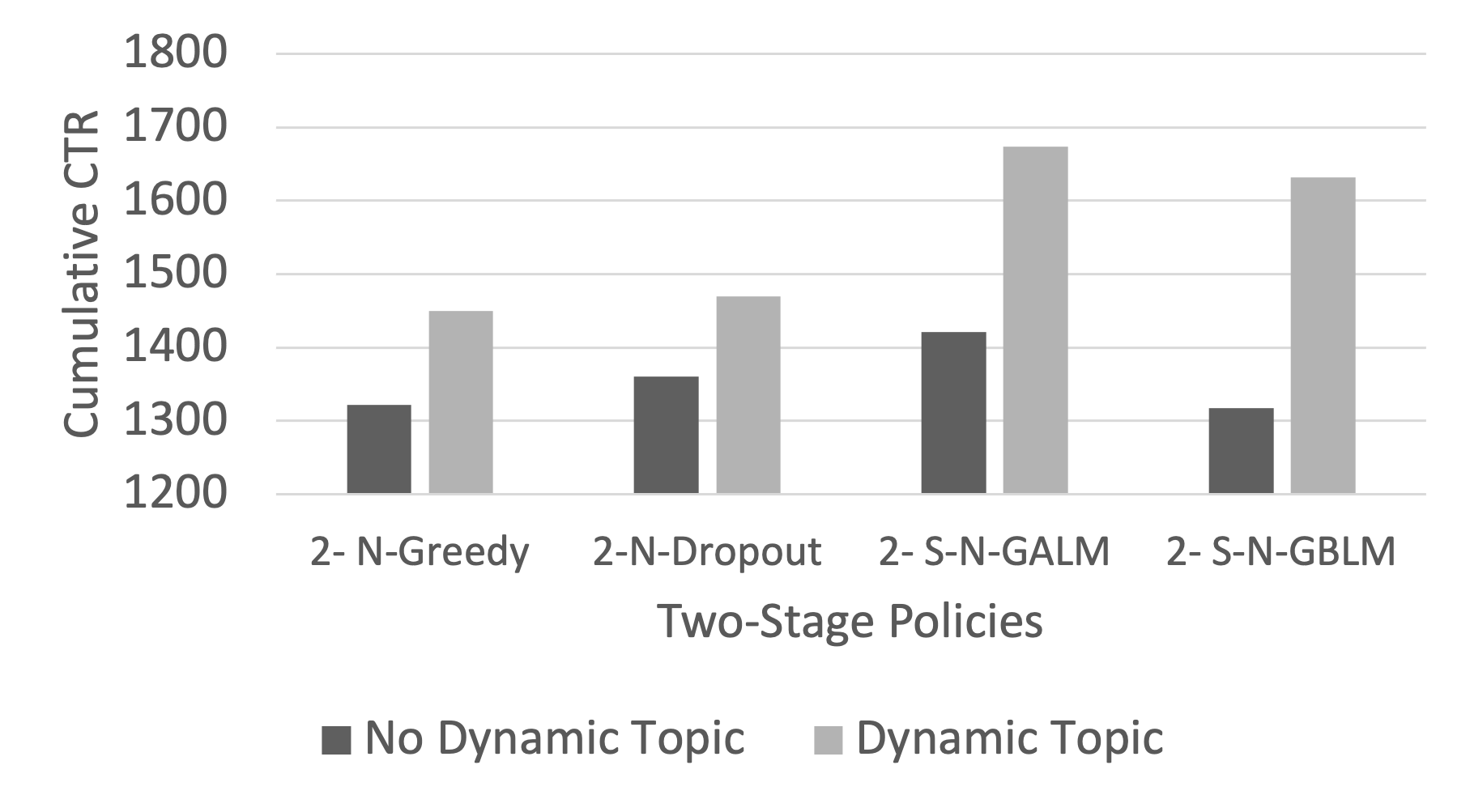}
    \end{subfigure}
    \caption{Results analysis and ablation study. All policies are UCB policies except the greedy policies.
    a) Improvement over random policy: the ratio of cumulative CTR of one-stage policies in Table \ref{table: one stage ctr} and cumulative CTR of random policy for different number of users.
    b) Influence of number of users: cumulative CTR ratio between different number of users for one-stage policies in Table \ref{table: one stage ctr}.
    c) Improvement of two-stage design: cumulative CTR ratio between two-stage and one-stage corresponding policies in Table \ref{table: two stage ctr}.
    d) Ablation study for dynamic topic comparison (2,000 iteration, recommendation size 5 and user size 100.). The prefix ``2-'' indicates two-stage policies. 
    }
    \label{fig: results analysis}
    \vspace{-12pt}
\end{figure}

\vspace{-3mm}
We show our observations from the main result tables in Section \ref{sec: main results} and the analysis in Figure \ref{fig: results analysis}.

\textbf{Two-tower neural representation improves the Performance:}
In Table \ref{table: one stage ctr}, compared with non-neural policies (first two rows), the neural network based policies (last 5 rows) has significant improvement. 
Figure \ref{subfig: Improvement over random policy} shows the cumulative CTR ratio between policies in Table \ref{table: one stage ctr} and random policies, which illustrates the improvement of neural based methods.
Particularly, the only difference between N-GLM-UCB and GLM-UCB is the N-GLM-UCB makes use of neural news item representation from two-tower model while GLM-UCB uses GLoVe directly. 
When there is enough samples for each user (e.g. 10 users, 200 samples each user), 
the cumulative CTR of N-GLM-UCB is 2.58 times of that of GLM-UCB.  
Our proposed policies further improves the cumulative CTR by also making use of neural user representation from two-tower model.
This shows the power of combining two-tower neural representations into the bandits recommendation framework. 

\textbf{Shared weights for bandit model improves CTR: }
In Table \ref{table: one stage ctr}, we can see the cumulative CTR for the disjoint policies like GLM-UCB and N-GLM-UCB drops dramatically when the number of users increases (i.e. number of samples per user decreases), which shows the disjoint models are hard to be scalable to the large user or item recommender system. 
This is because the disjoint policies need enough samples to learn the coefficients for each user, as discussed in Section \ref{sec: baseline policies}.
Both N-Greedy and N-Dropout-UCB outperform N-GLM-UCB on a large number of users, since both of the policies based on neural networks directly and no additional parameters need to be learnt for each user.
Our proposed policies, which extend from N-GLM-UCB to share parameters across different users or items, outperform disjoint policies as well. 
This is verified in Figure \ref{subfig: Influence of number of users}, where we show the cumulative CTR ratio between different number of users for the one-stage policies in Table \ref{table: one stage ctr}. Except the ratios of the disjoint policies (GLM-UCB, N-GLM-UCB) are much lower than $1$ (i.e. increase number of users, CTR drops), the ratio of other policies are around $1$.

\textbf{Two-stage outperforms one-stage:}
In Table \ref{table: two stage ctr}, two-stage policies outperform corresponding one-stage policies since the topic exploration scope in the news space under promising topics and save the computational budget. 
The exception is the two-stage Random policy, which is worse than one stage Random since selecting bad topics at the first stage would limit the news selection and lead to a lower click rate. This further shows the importance of a reasonable topic recommendation. 
We visualise the cumulative CTR ratio between two-stage and one-stage policies in Figure \ref{subfig: Improvement of two-stage design}, where we can see except the random policy (and N-Greedy, N-Dropout-UCB for $10$ recommendations), the ratios are above $1$. 
The improvements of $5$ recommendations and our proposed policies are higher than others.

\textbf{Proposed bandit policies outperform others: }
Both of our proposed policies (S-N-GBLM-UCB, S-N-GALM-UCB) have higher cumulative CTR compared other polices in Table \ref{table: one stage ctr} and \ref{table: two stage ctr}, which illustrates the effectiveness of the shared model and usage of the user representation from neural network with additive (S-N-GALM-UCB) and bilinear (S-N-GBLM-UCB) structure. 
With 2,000 iterations, S-N-GALM-UCB slightly outperforms S-N-GBLM-UCB. 
Although S-N-GBLM-UCB is able to capture user and item interaction in the generalised linear part, the larger parameter space needs more time and sample to learn. 

\textbf{Dynamic topic clustering improves CTR:}
We further show ablation study in Figure \ref{subfig: Ablation study for dynamic topic comparison}, where we test how our dynamic topic construction in Algorithm \ref{alg: Dynamic Topic Reconstruction} influences the performance of two-stage policies in Table \ref{table: two stage ctr}. 
We set the minimum reconstruct size $p = 1000$. 
We compare with the case where only the top topic is recommended.
To make the comparison fair, if the candidates news under the recommended topic is smaller than the computational budget, we randomly sample news from the whole news set to guarantee the number of news evaluated for two methods are the same.
We can observe that for all policies, using the dynamic topic construction improves the CTR significantly. 



\vspace{-2mm}
\section{Conclusion}
\label{sec:discussion}
\vspace{-2mm}
We consider the news recommendation task in the contextual bandits setting to balance the exploitation and exploration in the sequential decision-making process.
We propose a two-stage topic-news recommendation framework with dynamically generated topics, to increase computational efficiency in the large arm space.
We utilise the deep representation from a two-tower neural model and propose the generalised additive and bilinear upper confidence bound policies to generate uncertainties.
Empirical experiment on a large-scale news recommendation dataset shows our proposed two-stage framework is computationally efficient and our proposed policies outperforms baselines.


\newpage
\bibliography{cb4rec}
\bibliographystyle{plainnat}

\newpage
\appendix
\section*{Appendix} 

\begin{algorithm}[h]
  \caption{Dynamic Topic Reconstruction}
  \label{alg: Dynamic Topic Reconstruction}
  \textbf{Input:}{
  $l_{st}:$ topic list sorted according to topic acquisition score in non-increasing order, constructed topic group arms $\mathcal{S}^1, \dots, \mathcal{S}^m$,
   minimum reconstruction topic size $p$}.
  \begin{algorithmic}[1]
      \While{there exists at least one topic group arm has size smaller than $p$}
        \For{$v = 1$ to $m$}
            \If{$|\mathcal{S}^v| < p$}
                \State{$\mathcal{S}^v \leftarrow \mathcal{S}^v \cup l_{st}[0]; l_{st} = l_{st}/l_{st}[0]$}
            \EndIf
        \EndFor
      \EndWhile
  \end{algorithmic}
\textbf{Return} Reconstructed topic group arms $\mathcal{S}^1, \dots, \mathcal{S}^m$.
\end{algorithm}

\section{Supplementary Experiment Details}
\label{sec: appendix experiment details}

Our experiment is conducted in python 3.8 (with PyTorch 1.9). We run our experiments on 2 Titan V GPUs. We provide our code and instructions to reproduce our main experiments in supplementary materials, and provide more experiment details below.

\subsection{Simulated Rewards}
\paragraph*{Off-policy user feedback training}
In this part, we describe in details our training method to simulate user feedback based on a large-scale news recommendation dataset (MIND) \citep{wu_mind_2020}. We build the user feedback module upon the neural news recommendation with multi-head self-attention (NRMS) \citep{wu_nrms_2019}. Specifically, given a user $u \in \mathcal{U}$ and an news item $i \in \mathcal{A}$, NRMS builds a user encoder $f^u: \mathcal{U} \rightarrow \mathbb{R}^d$ and a news encoder $f^i: \mathcal{A} \rightarrow \mathbb{R}^d$, where the architectures for $f^u$ and $f^u$ are given in \citep[Figure~2]{wu_nrms_2019}. Given such encoders, the click probability score is computed by the inner product of the user representation vector and the news representation vector, i.e. $\hat{y}_{u,i} = f^u(u)^T f^i(i)$. 

To train such a click probability score model above, initially \citet{wu_nrms_2019} use softmax loss with negative sampling techniques. That is, for a given user, each news browsed by the user (regarded as a positive sample) is combined with $K$ randomly sampled news in the same impression but not clicked by the user (regarded as negative samples) to form a set of samples with the corresponding click probability scores $\hat{y}_i^+, \hat{y}^-_{i,1}, \ldots, \hat{y}^-_{i,K}$. 
The softmax score for the positive sample is then computed as 
\begin{align*}
    p_i = \frac{\exp(y^+_i)}{\sum_{j=1}^K \exp(y^-_{i,j})}. 
\end{align*}
The final loss function is the negative log-likelihood of all positive samples $S$: 
\begin{align*}
    \mathcal{L} = - \sum_{i \in S} \log p_i. 
\end{align*}

To simulate binary rewards, in our work, we instead using binary cross-entropy loss with negative sampling techniques. In particular, with the same notations above, the binary cross-entropy we used in our work is 
\begin{align*}
    \mathcal{L}_{BCE} := \sum_{i \in S} (bce(y^+_i) + \sum_{j=1}^K bce(y^{-}_{i,j})),
\end{align*}
where $bce(y^+_i) := - \log \textrm{sigmoid}(y^+_i)$ and $bce(y^-_{i,j}) : = - \log (1 - \textrm{sigmoid}(y^-_{i,j}))$. 

However, $\mathcal{L}_{BCE}$ is biased as samples (including both positive and negative samples) are not equally distributed, as the fixed dataset has been collected by some unknown behaviour policy, which is not necessarily a uniform sampling. To de-bias our initially proposed loss $\mathcal{L}_{BCE}$, we leverage an off-policy evaluation approach via Hájek estimator
\begin{align*}
    \mathcal{L}_{BCE}^{debiased} := \sum_{i \in S} \frac{\frac{bce(y^+_i)}{P(O_{u,i} = 1)} + \sum_{j=1}^K \frac{bce(y^-_{i,j})}{P(O_{u,j} = 1)}}{\frac{1}{P(O_{u,i} = 1)} + \sum_{j=1}^K \frac{1}{P(O_{u,j} = 1)} },
\end{align*}
where $O_{u,i} \in \{0,1\}$ is the random variable that indicates if the feedback is observed for a user-item pair $(u,i)$, and $u$ denotes the user associated with the positive sample $i$ in the current impression list (note that in each impression list in MIND dataset is associated with a unique user). In practice, we simply estimate $P(O_{u,i} = 1)$ using its empirical estimate directly from the dataset: 
\begin{align}
    \hat{P}(O_{u,i} = 1) = \frac{\# \text{of times } u \text{ sees } i }{\# \text{of times } u \text{ appears}}
\end{align}



\paragraph{Result} Our debiased training method described above produces a user feedback simulator with AUC score of $68.46\%$, higher than the reported AUC of $68.18\%$ of the original NRMS trained with the negative sampling techniques \citep[Table~2]{wu_empowering_2021}. We used $K = 4$ in our experiment.

\paragraph{Binary feedback simulation} Given our trained user feedback model above, we need to convert the click probability score $\hat{y}_{u,i}$ into a binary reward. For this, we first convert the click probability score $\hat{y}_{u,i}$ into a valid probability by applying the sigmoid function on the score. We then pick a threshold by maximizing $f$-score of the predicted probabilities over the entire dataset. As a result, we obtain the threshold value $THRES = 0.38414$. Then the simulated binary reward is $y_{u,i} := \mathbbm{1}\{\hat{y}_{u,i} \geq THRES\}$. Such threshold approach gives a deterministic binary reward. In practice, however, the user feedback can be stochastic. For example, given a fixed user and fixed news, the user might not click on that news when (s)he sees the news for the first time but not for later times when (s)he sees the news again as this time his/her preference might have changed. To model such user preference uncertainty, we simply flip the value of the deterministic reward $y_{u,i} = \mathbbm{1}\{\hat{y}_{u,i} \geq THRES\}$ with some probability $p$. This flipping reward is our modelling choice rather than a data-driven choice as it is difficult to infer a user's preference uncertainty from a fixed dataset. In our experiment, we used $p = 0.1$.

\subsection{NRMS Neural Model}

\paragraph{News Item Neural Model}

For news items, we follow the NRMS model \citep{wu_nrms_2019}.
It contains \textit{news encoder} and \textit{user encoder}.
The news encoder learns news representation from news titles, which contains a word embedding layer, word-level multi-head self-attention and additive work attention network. 
The user encoder learns user representation from their browsed news, which contains news-level multi-head self-attention and additive news attention network. 
We follow the hyperparameter settings in \citep{wu_nrms_2019} and change the news representation dimension to 64. 

\paragraph{Topic Neural Model}

We use the same architecture and hyperparameter settings of \textit{news encoder} and \textit{user encoder} as in the News Item Neural Model to get user representation. For each topic, we randomly initialise a vector with the same dimension of the user representation.
The topic encoder takes the topic name as input and contains a word embedding layer and a multi-layer perceptron.
We use dot product between the user representation and the topic representation to get user interest in topics in stage one and train the model with binary cross-entropy. 
To balance the positive and negative samples, we further adopt the negative sampling approach \citep{wu_nrms_2019} with positive and negative sample ratio as 1. 

\subsection{Simulation Settings}
We specify the additional parameter setting for simulation which has not been specified in the main paper.
We update item neural models every 100 iterations, topic neural models every 50 iterations, topic and item generalised linear models every iteration (if there exist clicks from recommendations). 
We inference Monte-Carlo dropout 5 times and dropout rate is set to be 0.2.
Dropout is applied in the news encoder after the word embedding layer and multi-head attention layer. 
Since our user representation is based on the clicked news representation, the dropout uncertainty includes both the news and user uncertainty.
We set the minimum topic construction size as 1,000. 
We train the generalised linear models by gradient descent and select the learning rate as 0.01 (for bilinear learning rate is 0.001).
Except specified, we set the UCB parameter $\beta = 0.1$.

\subsection{Additional experiments and observations}

\textbf{Small recommendation size has low CTR: }
In Table \ref{table: two stage ctr}, we can observe for all policies, the cumulative CTR for recommendation size 1 is much smaller than those of recommendation sizes 5 and 10.
The reason is that the small recommendation size restricts the number of feedback the system can get and thus with the same number of iterations, the parameter learning is slow for both neural and generalised linear models with recommendation size 1.
Additionally, we can observe that the N-Greedy and N-Dropout-UCB based policies have a large variance for the top-1 recommendation, while our proposed policies perform more stable in this case. 

\begin{figure}[t!]
    \centering
    \includegraphics[scale=0.3]{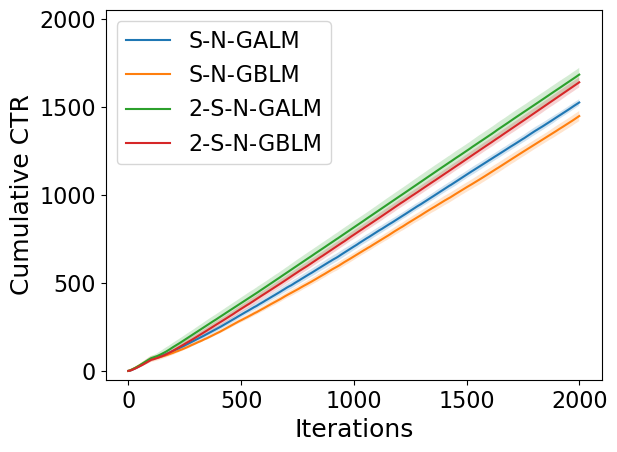}
    \hspace{0.5cm}
    \includegraphics[scale=0.3]{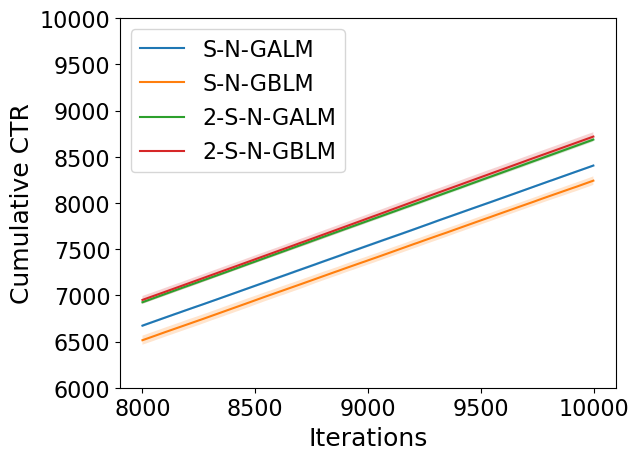}
    \caption{Cumulative CTR for proposed one and two stage algorithms for T = 10,000. The left and right show the first and last 2,000 iterations respectively. The experiment is conduct with 5 trials, recommendation size 5 and user size 100.
    }
    \label{fig: largeT}
\end{figure}

\textbf{Cumulative CTR for large iterations:}
In Table \ref{table: one stage ctr} and \ref{table: two stage ctr}, we can see additive based policies have higher cumulative CTR than bilinear based models for 2,000 iterations. We further verify our interpretation about bilinear policies take more time to train and need more samples to learn. 
We compare the cumulative CTR curves for our proposed policies in one stage and two stages with 10,000 iterations. We show the curves for first 2,000 iterations (left) and last 2,000 iterations (right) in Figure \ref{fig: largeT}.
We can see on the left, for both one and two stages, GALM outperforms GBLM for small iterations, while on the right, 2-S-N-GALM and 2-S-N-GBLM have similar performance (2-S-N-GBLM is lightly better), and S-N-GALM is still better than S-N-GBLM.

\begin{figure}[t!]
    \centering
    \includegraphics[scale=0.45]{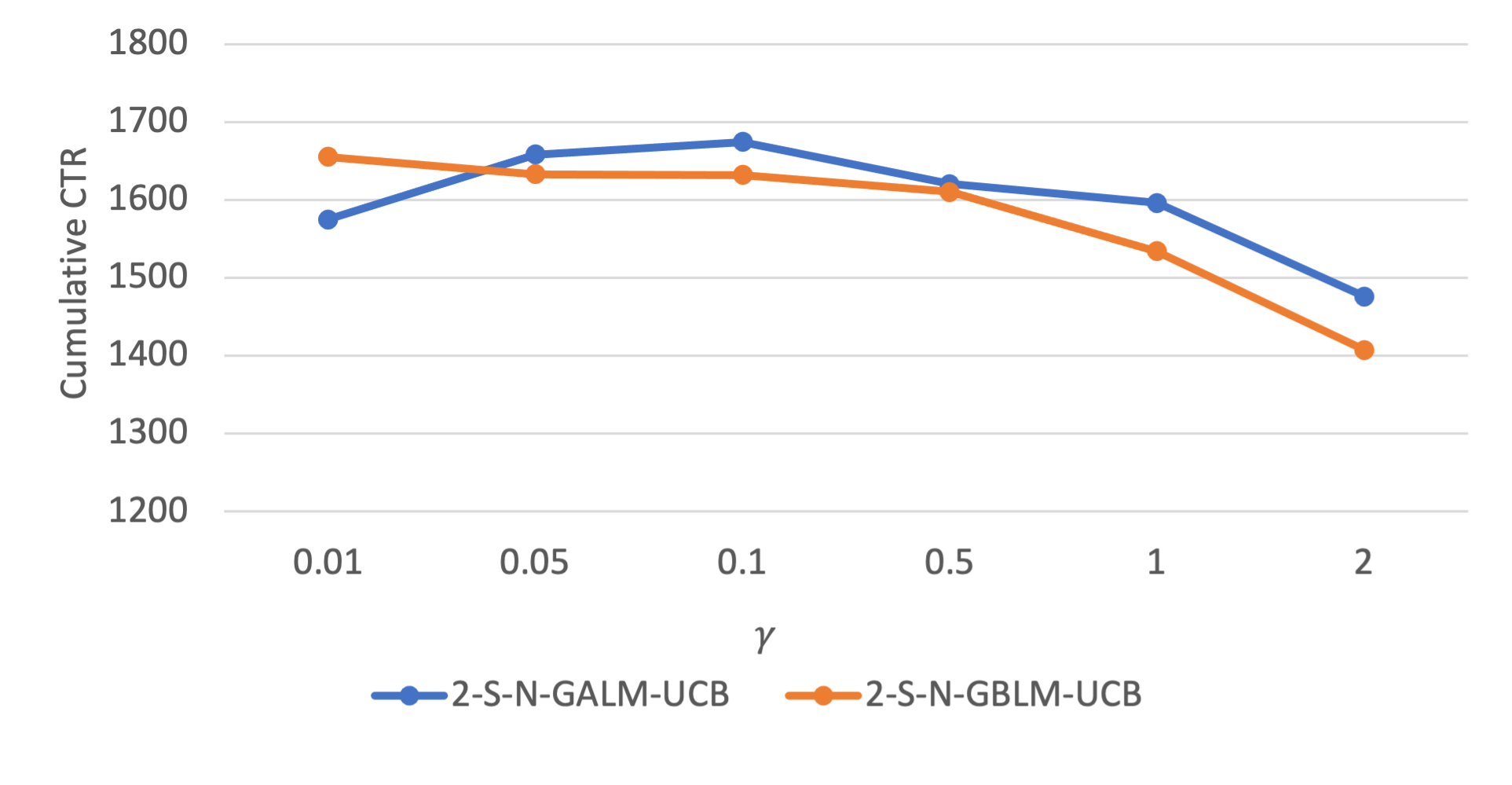}
    \caption{UCB Hyperarameter $\beta$ tuning. ``2-"" represents two-stage policies. The experiment is conducted with 2,000 iteration, recommendation size 5 and user size 100. 
    }
    \label{fig: beta comparison}
\end{figure}

\textbf{Hyperparameter tuning:}
In Figure \ref{fig: beta comparison}, we show how changing the exploitation-exploration balance parameter $\beta$ influences the cumulative CTR for our two-stage proposed algorithms, where the choice is $\beta$ is the same for both the topic and item stage. 
We can observe that for both of the policies, relatively small $\beta$ gives a high CTR.
In particular, a large $\beta$ (e.g. $\beta = 2$) will lead to a performance drop since it involves too much exploration.


\end{document}